\documentclass[aps,prl,twocolumn,showpacs,superscriptaddress,letterpaper]{revtex4}
\usepackage{graphicx,epsfig,amsmath,amssymb,amsfonts,mathrsfs,psfrag,color}
\usepackage{bm,txfonts,mathrsfs,mathtools,relsize}

\newcommand{\round}[1]{\left( #1 \right)}
\renewcommand{\square}[1]{\left[ #1 \right]}
\newcommand{\ang}[1]{\left\langle #1 \right\rangle}

\begin{document} 

\title{Squeezed noise due to two-level system defects in superconducting resonator circuits}
\author{So Takei}
\affiliation{Condensed Matter Theory Center, Department of Physics, The University of Maryland College Park, MD 20742}
\author{Victor M. Galitski}
\affiliation{Condensed Matter Theory Center, Department of Physics, The University of Maryland College Park, MD 20742}
\affiliation{Joint Quantum Institute, Department of Physics, The University of Maryland College Park, MD 20742}
\author{Kevin D. Osborn}
\affiliation{Laboratory for Physical Sciences, College Park, MD 20740}
\date{\today}

\begin{abstract}
Motivated by recent surprising experimental results for the noise output of superconducting microfabricated resonators 
used in quantum computing applications and astronomy, we develop a fully quantum theoretical model to describe quantum 
dynamics of these circuits. Building on theoretical techniques from quantum optics, we calculate the 
noise in the output voltage due to two-level system (TLS) defects. The theory predicts squeezing 
for the noise in the amplitude quadrature with respect to the input noise, which qualitatively reproduces the noise ellipse 
observed in experiment. We show that noise enhancement along the phase direction persists for pump frequencies away from
resonance. Our results also suggest that intrinsic TLS fluctuations must be incorporated in the model in order to describe 
the experimentally observed dependence of the phase noise on input power.
\end{abstract}

\maketitle

Amorphous dielectrics contain weakly-coupled two-level system (TLS) defects \cite{philippsrev,book} that are known to modify 
the complex permittivity at low temperatures, including the imaginary part which is responsible for loss \cite{schickfus}. 
In Josephson qubits the purposeful amorphous dielectrics were found to contain TLSs which leads to decoherence \cite{martinisetal}. 
Similarly, the loss in superconducting resonators is generally attributed to TLS defects which can arise from the native oxides on 
the superconducting leads \cite{wangetal} or deposited amorphous dielectrics \cite{simmondsetal,kevins}. Superconducting resonators 
are used in applications of quantum computing \cite{qubits} and as single photon detectors for astronomy, where noise can limit the 
detection performance \cite{thesis} and may be attributable to TLSs \cite{gaokumar}. Surprisingly, recent studies have revealed that 
the amplitude noise quadrature of resonators is limited by the experimental sensitivity near the vacuum noise limit \cite{gaoarxiv}. 
In contrast, noise in the phase quadrature is relatively large and decreases as the square root of the measurement power 
\cite{gaoetal}. A semi-empirical model was used to describe this phase noise saturation, but a quantitative model of phase-amplitude
noise asymmetry is still lacking \cite{gaoetal2}.

Previous related noise studies have considered the effects of TLS noise on superconducting qubits \cite{yunoise}, and noise
squeezing in various nonlinear systems including Josephson junction 
parametric amplifiers \cite{yurkeetal,beltranetal}, micro-cantilevers \cite{rugar}, and nanomechanical resonators coupled to a 
Cooper-pair box \cite{suhetal}. Atoms have also been used for four-wave mixing experiments to squeeze optical cavity states 
\cite{slusheretal}.

\begin{figure}[t]
\centering
\includegraphics*[scale=0.29]{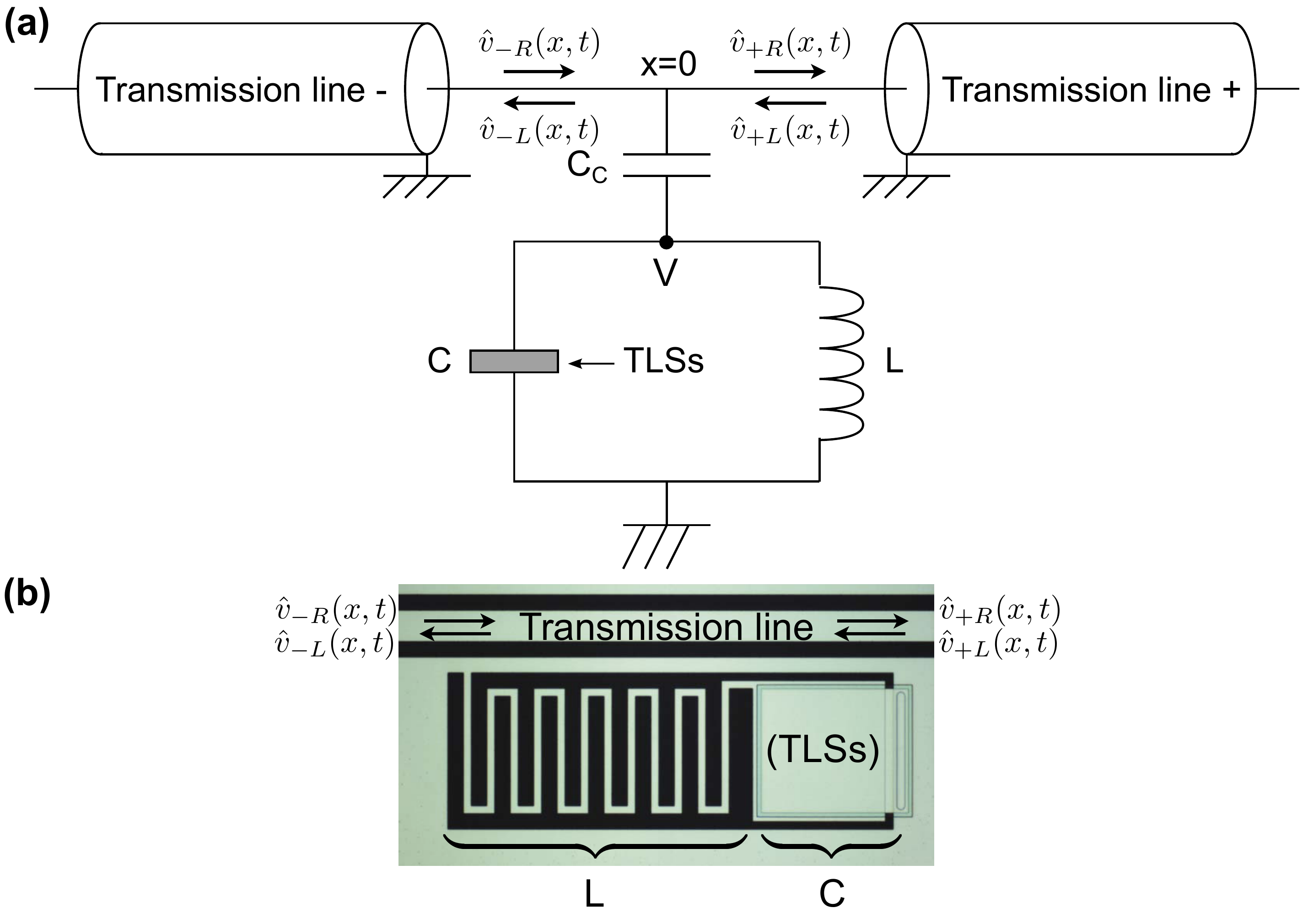}
\caption{\label{fig:circuit} (a) A circuit diagram of the considered setup. (b) Actual image (taken from Ref. \onlinecite{kevins}) 
of a notch-type aluminum LC resonator which can be modelled using (a). The components are labelled to clarify its correspondence 
to circuit in (a).}
\end{figure}

In this letter, we develop a theory for noise due to TLS defects in a superconducting resonator circuit. We compute the noise
in the transmitted voltage for the system shown in Fig. \ref{fig:circuit}(a) assuming that the lone source of stochastic fluctuations are 
the quadrature-independent white noise in the incident voltages on the transmission lines. The theory predicts that squeezing in the 
transmitted noise generally occurs even if the dynamics of the internal resonator, including the TLSs, is fully
deterministic. The results imply that intrinsic stochastic dynamics of TLSs is not necessary in explaining the observed
squeezing in the transmitted voltage noise. Our work shows a similarity between 
noise squeezing from TLS defects in these devices \cite{gaoarxiv,gaoetal} and squeezing of coherent light due to two-level 
atoms in quantum optics \cite{lugiatorev}. We also compute noise power for off-resonant pump frequencies and confirm that 
enhancement in the phase noise and squeezing in the amplitude noise persists away from resonance.

We have two identical semi-infinite 1D transmission lines, labeled $-$ and $+$, which are attached to each other end-to-end 
at $x=0$ (see Fig. \ref{fig:circuit}(a)). 
They are coupled to the internal superconducting LC resonator at $x=0$ through the coupling capacitor $C_c$. The 
capacitor and inductor in the LC circuit have capacitance $C$ and inductance $L$. Here, we take $C_c\ll C$, and 
consider the case where TLS defects reside only inside the  dielectric of capacitor $C$. A microwave source (not shown) sends in 
a pump signal from $x=-\infty$, and we focus on the transmitted voltage noise on transmission line +. 

The Hamiltonian for the full system has three terms, $\hat H=\hat H_0+\hat H_{\rm TLS}+
\hat H_{{\rm TLS}-R}$. $\hat H_0=\sum_{m=\{+,-\}}\hat H_m+\hat H_R$ models the two transmission lines and the 
internal resonator,
\begin{eqnarray}
\centering
\label{li}
\hat H_m&=&\int dx\, \Theta(mx)\square{\frac{\hat P_m^2(x)}{2\ell}+\frac{\round{\nabla_x\hat Q_m(x)}^2}{2c}}\\
\label{lr}
\hat H_R&=&\frac{\hat P_L^2}{2L}+\frac{\hat Q_0^2}{2C_c}+\frac{(\hat Q_0-\hat Q_L)^2}{2C},
\end{eqnarray}
where $m=\{+,-\}$ labels the left ($-$) and right ($+$) transmission lines, and $\Theta(x)$ is the unit step function. 
$\hat Q_m(x)$ is the operator for total charge residing to the right of $x$ on transmission line $m$, $\hat Q_L$ is the 
operator for total charge that has flowed through the inductor, and $\hat Q_0=(\hat Q_--\hat Q_+)|_{x=0}$ denotes 
the total charge operator on capacitor $C_c$. $\hat P_m$ and $\hat P_L$ are the conjugate momenta for $\hat Q_m$ 
and $\hat Q_L$, respectively, and they satisfy $[\hat Q_m(x),\hat P_{m'}(x')]=i\hbar\delta_{mm'}\delta(x-x')$ and 
$[\hat Q_L,\hat P_L]=i\hbar$, where $m,m'=\{+,-\}$. 
The transmission lines are modeled as conductors with inductance per unit length $\ell$, capacitance to ground 
per unit length $c$, and characteristic impedance $Z=\sqrt{\ell/c}$ \cite{yurke,yurkedenker}. We assume we have
$N$ identical and independent TLSs, all with asymmetry energy $\Delta_A$ and 
tunneling energy $\Delta_0$. The TLS Hamiltonian  is then given by $\hat H_{\rm TLS}=\Delta_A\hat S_z+\Delta_0\hat S_x$, 
where $\hat S_i=\sum_{\alpha=1}^N\hat s_{i\alpha}$ is the collective spin operator which represents the $N$ TLSs, and $\hat s_{i\alpha}
=\sigma_i/2$ with the usual Pauli matrices $\sigma_i$. The components of the collective spin operator obey
$[\hat S_i,\hat S_j]=i\epsilon_{ijk}\hat S_k$. After diagonalization $\hat H_{\rm TLS}=E\hat S_z$ where 
$E=[\Delta_A^2+\Delta_0^2]^{1/2}$. Each TLS interacts with the uniform electric field inside the capacitor through its electric dipole 
moment $\bf p$. We assume the dipoles fluctuate by making $180^\circ$ flips between parallel and anti-parallel
orientations with respect to the field. The field affects the asymmetry energy and here we ignore the relatively small changes to the 
tunnel barrier, similar to other treatments of TLSs derived from the tunneling model \cite{philippsrev}. 
In the diagonalized basis the interaction between the TLSs and the field can then be written as
$\hat H_{{\rm TLS}-R}=g\hbar[\hat S_z\cot2\xi+(\hat S^++\hat S^-)/2](\hat Q_0-\hat Q_L)$,
where the coupling constant $g=2|{\bf p}|\sin 2\xi/\hbar Cd$, $d$ is the separation between the plates of the 
capacitor, $\xi$ is defined via $\tan2\xi=\Delta_0/\Delta_A$, and $\hat S^\pm=\hat S_x\pm i\hat S_y$. We use $|{\bf p}|=1$ Debye, 
which is comparable to the dipole moment sizes observed for TLSs in amorphous SiO$_2$ at microwave frequencies and OH rotors in 
AlO$_x$ \cite{musgrave}. 

For $x\ne 0$, $\hat Q_m(x,t)$ obey the massless scalar Klein-Gordon equation, 
and the solution can be written as a sum of right- ($R$) and left-propagating ($L$) components, 
i.e. $\hat Q_m(x,t)=\hat Q_{mR}(x,t)+\hat Q_{mL}(x,t)$, where
\begin{equation}
\label{chargefieldexp}
\hat Q_{m\{R,L\}}(x,t)=\int_0^\infty\frac{d\omega}{2\pi}\sqrt{\frac{\Omega_r}{\omega}}
(\hat q_{m\{R,L\}}^\dag(\omega)e^{i\omega(t\mp x/v)}+h.c.),
\end{equation}
$v=(\ell c)^{-1/2}$ is the velocity of wave propagation and $\Omega_r$ is the resonance frequency of the loaded LC circuit. 
The charge operators obey $[\hat q_{m\tau}(\omega),\hat q^\dag_{m'\tau'}(\omega')]=(\pi\hbar/\Omega_rZ)\delta_{mm'}\delta_{\tau\tau'}
\delta(\omega-\omega')$, where $\tau,\tau'=\{R,L\}$. We also introduce the ladder operator for the inductor charge via
$\hat Q_L=\hat q_L+ \hat q_L^\dag$, where $[\hat q_L,\hat q_L^\dag]=\hbar/2\omega_0L$ and $\omega_0=(LC)^{-1/2}$.

At $x=0$, the quantum Maxwell-Bloch equations for our problem read
\begin{eqnarray}
\label{eom1na}
\left.\nabla_x(\hat Q_{-R}+\hat Q_{-L})\right|_{x=0^-}&=&
\left.\nabla_x(\hat Q_{+R}+\hat Q_{+L})\right|_{x=0^+}\\
\label{eom2na}
\left.\frac{\nabla_x(\hat Q_{+R}+\hat Q_{+L})}{c}\right|_{x=0^+}&=&  
-\round{\frac{\hat Q_0}{C_c}+L\ddot {\hat Q}_L}\\
\label{eom3na}
LC\ddot {\hat Q}_L-(\hat Q_0-\hat Q_L)&=&Cg\hbar(\hat S_z\cot2\xi+(\hat S^++\hat S^-)/2)\\
\dot{\hat S}^-+[i(E/\hbar)+\Gamma_2]\hat S^-&=&ig(\hat S_z-\cot2\xi\hat S^-)(\hat Q_0-\hat Q_L)\\
\label{eom5na}
\dot{\hat S}_z+\Gamma_1(\hat S_z-S_z^0)&=&(ig/2)(\hat S^--\hat S^+)(\hat Q_0-\hat Q_L).
\end{eqnarray}
Here, we have introduced phenomenological longitudinal and transverse TLS relaxation rates, $\Gamma_1=T_1^{-1}$ and 
$\Gamma_2=T_2^{-1}$, and $\hat Q_0=[\left.\hat Q_{-R}+\hat Q_{-L}-\hat Q_{+R}-\hat Q_{+L}]\right|_{x=0}$. 
$S_z^0=-(N/2)\tanh(E/2k_BT)$ is the equilibrium expectation value for $\hat S_z$ when the TLSs are decoupled 
from the fields. Assuming we are in the low-temperature regime, i.e. $k_BT \ll E$, we take $S_z^0\approx -N/2$.
The first three equations above are quantum generalizations of the Kirchhoff voltage law, and the last two equations 
are quantum Bloch equations describing the coupling of the TLSs to the photon fields. 

We now perform a series of standard approximations which maps the above equations to a form 
often seen in quantum optics literature describing an ensemble of two-level atoms in an optical cavity \cite{lugiatorev}. 
Since we have a high-$Q$ resonator we perform the Markov approximation, where we replace $\sqrt{\omega}$ 
by $\sqrt{\Omega_r}$ in the mode expansion Eq. (\ref{chargefieldexp}) and extend the lower limit of the integral to negative 
infinity. We now go to a frame rotating at the pump frequency $\Omega$, and use Greek letters to define operators in the 
rotating frame, i.e. $\hat\theta_{m\tau}(t)=\hat q_{m\tau}(t)e^{i\Omega t}$, $\hat\theta_L(t)= 
\hat q_L(t)e^{i\Omega t}$, $\hat\Sigma^\pm(t)=\hat S^\pm(t)e^{\mp i\Omega t}$, and $\hat\Sigma_z(t)=\hat S_z(t)$. 
Performing the slowly-varying envelope approximation on the $\ddot Q_L(t)$ term and with the usual rotating-wave 
approximation, Eqs. (\ref{eom1na})-(\ref{eom5na}) can then be rewritten as 
\begin{eqnarray}
\label{eom1a}
(\partial_t-i\Omega)\hat\theta_1&=&0\\
\label{eom2a}
ZC_c(\partial_t-i\Omega)(\hat\theta_{+R}-\hat\theta_{+L})&=&\hat\theta_{0}
-LC_c\round{\Omega^2+2i\Omega\partial_t}\hat\theta_L\\
\label{eom3a}
-LC\round{\Omega^2+2i\Omega\partial_t}\hat\theta_L&=& 
\hat\theta_{0}-\hat\theta_L+(Cg\hbar/2)\hat\Sigma^-\\
(\partial_t+i\Delta+\Gamma_2)\hat\Sigma^-&=&ig\hat\Sigma_z(\hat\theta_{0}-\hat\theta_L)\\
\label{eom5a}
2\partial_t\hat\Sigma_z+2\Gamma_1(\hat\Sigma_z-\Sigma_z^0)&=&
ig[\hat\Sigma^-(\hat\theta_0^\dag-\hat\theta_L^\dag)-h.c.],
\end{eqnarray}
where $\Delta=E/\hbar-\Omega$ is the TLS detuning, and
$\hat\theta_{\{0,1\}}=\hat\theta_{-R}\pm\hat\theta_{-L}-\hat\theta_{+R}\mp\hat\theta_{+L}$.
We note that the voltage fields propagating along the transmission lines relate to the charge
fields $\hat\theta_{m\tau}$ through $\hat v_{m\tau}(t)=-c^{-1}\nabla_x\hat\theta_{m\tau}(x,t)|_{x=0}$.

We solve Eqs. (\ref{eom1a})-(\ref{eom5a}) for $N\gg 1$ where the dynamics of the quantum system can be
described semi-classically \cite{altlandetal}. We treat each field $\phi(t)$ as a $c$-number and assume it has a steady-state 
mean part $\bar\phi$ and a fluctuating part $\delta\phi(t)$ which models the noise arising from the incident voltage on the
transmission lines. Assuming small fluctuations, we linearize Eqs. (\ref{eom1a})-(\ref{eom5a}) with respect to these 
fluctuations about the steady-state value.

\begin{figure}[t]
\centering
\includegraphics*[scale=0.48]{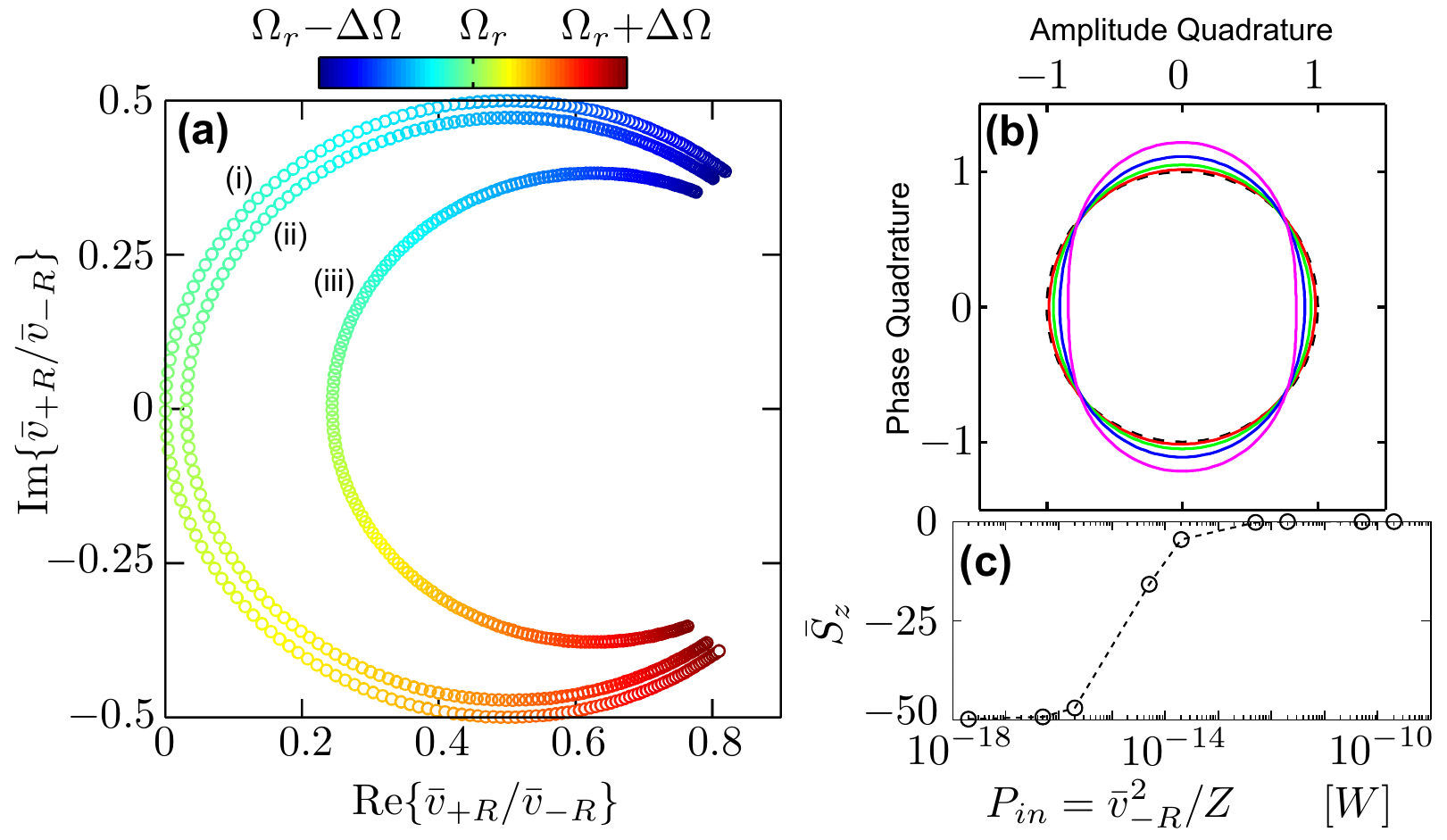}
\caption{\label{fig:MFres} (a) Steady-state solution for the transmission with $N=100$,
and (i) $\bar v_{-R}=10\mu$V; (ii) $\bar v_{-R}=1\mu$V; and (iii) $\bar v_{-R}=0.05\mu$V. Here,
$\Delta\Omega\approx 0.0003\omega_0$. (b) Noise
ellipses for $\bar v_{-R}=2\mu$V, $\Omega=\Omega_r$, and $N=100$ (red), $300$ (green), $600$ (blue), and $1000$ (magenta).
The dashed line is quadrature-independent noise
with no TLSs. (c) Plot of population imbalance, $\bar S_z$, as a function of input power for $N=100$.}
\end{figure}
The steady-state solution to the transmission amplitude $\bar v_{+R}/\bar v_{-R}$ is plotted in Fig. \ref{fig:MFres}(a) 
for real $\bar v_{-R}$ (input voltage amplitude) near the resonance frequency $\Omega_r$. Blue points correspond to 
pump frequencies below $\Omega_r$ and red points to those above $\Omega_r$.
The solutions are plotted for three different values of $\bar v_{-R}$, and the parameters used are $C_c=0.01$pF, $C=0.3$pF, 
$Z=50\Omega$, $f_0=\omega_0/2\pi=6$ GHz, $T_1=300$ns, $T_2=30$ns, $d\approx 70$nm and $N=100$. We also assume the TLSs 
to be in resonance with the LC resonator and fix the TLS energy to $E=\Omega_r$ throughout and take $\Delta_A=\Delta_0$. 
We then find $\Omega_r\approx 0.9837\omega_0$. In
Fig. \ref{fig:MFres}(c), $\bar S_z$ (TLS population imbalance) is plotted as a function of the input power, $P_{in}$. As $P_{in}$ 
increases $\bar S_z$ approaches zero signifying TLS saturation. We see that saturation occurs for 
$P_{in}\gtrsim 10^{-13}$W. Fig. \ref{fig:MFres}(a) shows that the 
transmission deviates very little from the defect-free (no TLSs) limit for saturated TLSs. This is expected since only a small fraction 
of the total energy is stored in the TLSs when they are saturated.

We now move on to obtaining the noise in the transmitted voltage. 
Solving the linearized equations the transmitted charge fluctuations can be written in terms of the two input fluctuations as
\begin{equation}
\centering
\label{tfluc}
{\bf A}_{+R}(\omega)\delta\Theta_{+R}(\omega)={\bf A}_{-R}(\omega)\delta\Theta_{-R}(\omega)
+{\bf A}_{+L}(\omega)\delta\Theta_{+L}(\omega),
\end{equation}
where $\omega$ is the frequency deviation measured from the pump frequency $\Omega$, 
$\delta\Theta_{m\tau}(\omega)=\round{\delta\theta_{m\tau}(\omega),\delta\theta_{m\tau}^*(-\omega)}^T$, 
and the elements of the coefficient matrices ${\bf A}_{m\tau}(\omega)$ are given in the Supplementary Material.
To study the quadrature-dependence of the noise in the transmitted voltage we introduce generalized 
voltage fluctuation variables
$\delta v_{m\tau}(\omega,\varphi)=[e^{-i\varphi}\delta v_{m\tau}(\omega)+e^{i\varphi}\delta v_{m\tau}^*(-\omega)]/2$,
where $\varphi$ is the quadrature angle measured with respect to the real axis. We focus on the
symmetrized correlator of these fluctuations, i.e. $S_{+R}(\omega,\omega',\varphi)=\ang{\{\delta v_{+R}(\omega,\varphi),\delta v_{+R}^*
(\omega',\varphi)\}}$, where $\{A,B\}=AB+BA$. Assuming that the incoming fluctuations, $\delta v_{-R}(t)$ and $\delta v_{+L}(t)$, are 
characterized by a quadrature-independent white noise spectrum $s_0$ (for thermalized transmission lines it is 
$s_0^{\rm th}=2\pi\hbar Z\Omega_r\coth(\hbar\Omega_r/2k_BT)$), we have
\begin{equation}
\centering
\label{noisein}
\ang{\{\delta v_{\{-R.+L\}}(\omega,\varphi),\delta v_{\{-R,+L\}}^*(\omega',\varphi)\}}=s_0\delta(\omega-\omega').
\end{equation} 
Using Eqs. (\ref{tfluc}) and (\ref{noisein}) together with $\delta v_{+R}(\omega,\varphi)\approx-2iZ\Omega_r
\delta\theta_{+R}(\omega,\varphi)$, $S_{+R}(\omega,\omega',\varphi)=s_{+R}(\omega,\varphi)\delta(\omega-\omega')$ 
can be straightforwardly obtained. (see Supplementary Material for details). In the following, we focus on the resulting spectral 
density of the noise, $s_{+R}(\omega,\varphi)$.

\begin{figure}[t]
\centering
\includegraphics[scale=0.57]{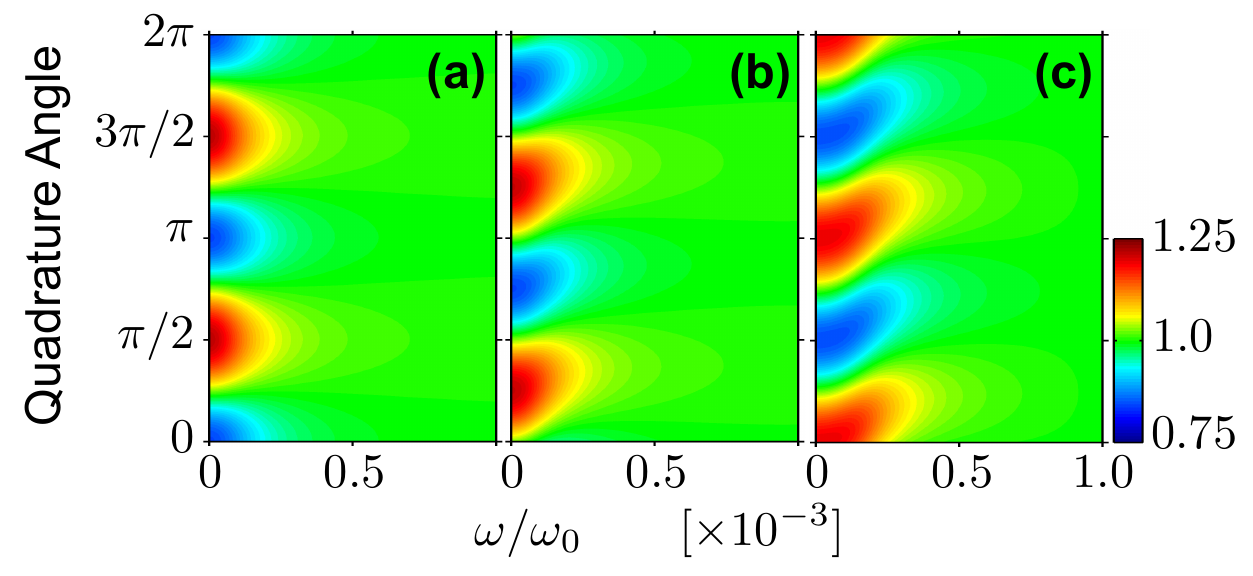}
\caption{\label{fig:noises2} Colour intensity plot of the normalized noise power $s_{+R}(\omega,\varphi)/s_0$ 
on the ($\omega$, $\varphi$)-plane for pump frequencies on and away from resonance, $\bar v_{-R}=2\mu$V and 
$N=1000$. Plots correspond to noise 
centred around (a) west; (b) northwest; and (c) north points, where north is defined as the top point on the
resonance circle.}
\end{figure}
We first focus on the on-resonance case (where $\mbox{Im}\{\bar v_{+R}\}=0$). 
Fig. \ref{fig:MFres}(b) plots noise ellipses centred at this resonance point for $\bar v_{-R}=2\mu$V and various $N$. 
Here, we are plotting the $\omega=0$ component of the noise spectral density normalized by $s_0$. 
The noise ellipse is defined such that every vector from the origin to a point on the ellipse makes an angle 
$\varphi$ (quadrature angle) with respect to the positive real axis and has length $s_{+R}(\omega=0,\varphi)/s_0$. 
In the defect-free limit, the noise ellipse is circular with radius 1 (the dashed line), which shows that in the absence of
non-linearity the transmitted noise remains quadrature-independent. In the presence of TLSs, we obtain \textit{squeezing} 
where the noise along amplitude (phase) quadrature is reduced below (enhanced above) the 
noise of the incoming fluctuations. Eccentricity of the noise ellipse increases as the number of TLSs is increased.

Fig. \ref{fig:noises2} is a colour intensity plot of the normalized noise power $s_{+R}(\omega,\varphi)/s_0$ 
on the $(\omega,\varphi)$-plane for on- and off-resonance pump frequencies. Here, $\bar v_{-R}=2\mu$V and $N=1000$ are 
both fixed. We now use 
the convention where north corresponds to the point at the top of the resonance circle in Fig. \ref{fig:MFres}(a), and west
to the left-most point on the circle and so on.
Then, the three plots correspond to the noise centred around (a) west; (b) northwest; and (c) north. 
The results show that the major axis of the noise ellipse is always along the phase direction.
Our results are consistent with experiments, where fluctuations are also primarily observed in the direction tangent to the 
resonance circle \cite{gaoetal}.

\begin{figure}[t]
\centering
\includegraphics[scale=0.45]{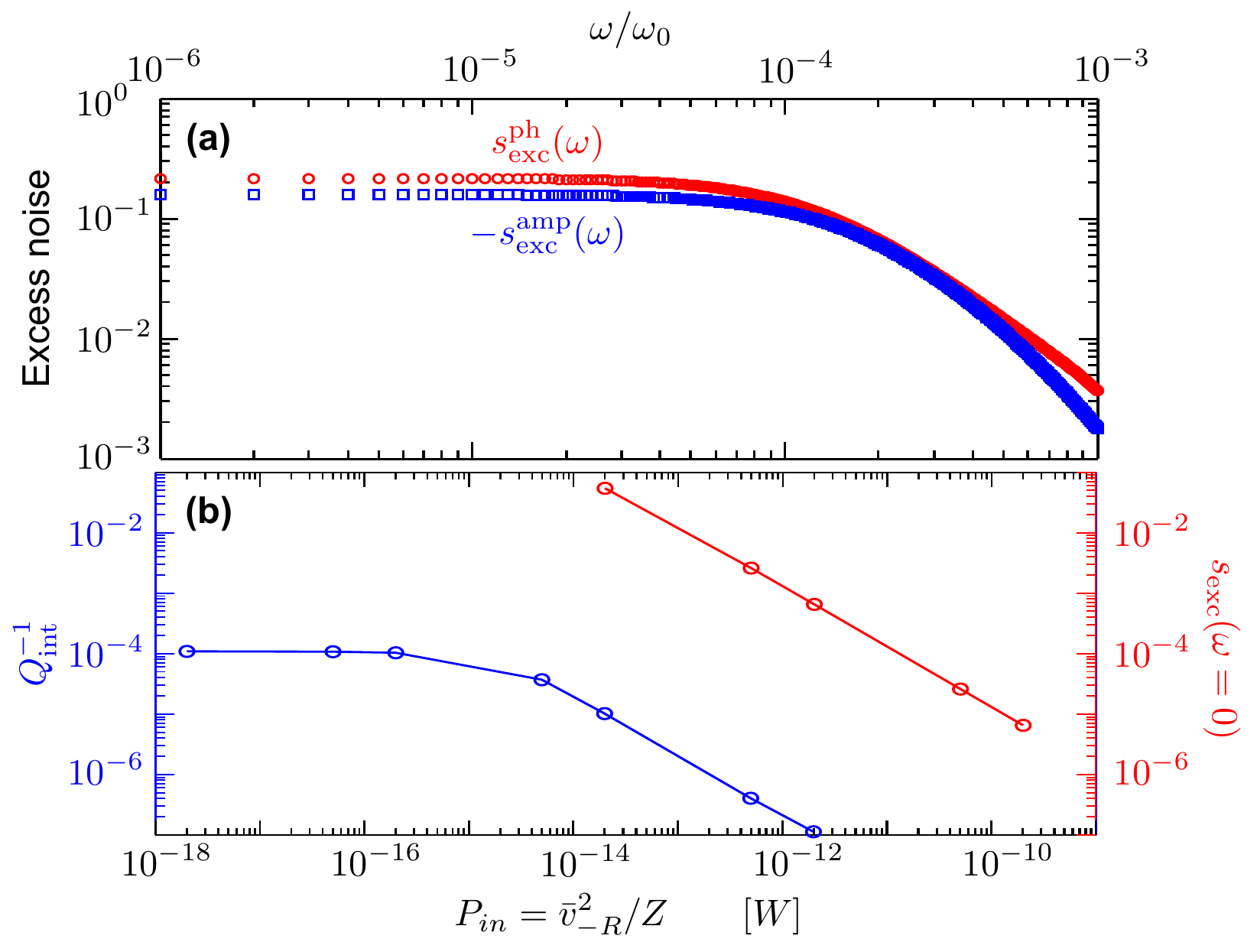}
\caption{\label{fig:powerdep} (a) Frequency-dependence of the excess phase noise power and negative of the excess
amplitude noise power for $\bar v_{-R}=2\mu$V and $N=1000$. Roll-off at $\approx\omega/\omega_0=10^{-4}$ can be seen.
(b) Normalized excess phase noise at $\omega=0$ and inverse of the internal quality factor as a function of input power 
($\Omega=\Omega_r$).}
\end{figure}

In Fig. \ref{fig:powerdep}(a) we plot the normalized excess phase noise power, 
$s^{\rm ph}_{\rm exc}(\omega):=[s_{+R}(\omega,\varphi=\pi/2)-s_0]/s_0$ and negative of the normalized excess
amplitude noise, $-s^{\rm amp}_{\rm exc}(\omega):=-[s_{+R}(\omega,\varphi=0)-s_0]/s_0$, as a function of frequency
deviation away from resonance, $\omega$. Here we use $\bar v_{-R}=2\mu$V and $N=1000$. 
Both noises roll off at the resonator bandwidth $\omega_{\rm roll}/\omega_0\approx 10^{-4}$. From the steady-state solution
the quality factor of the resonator is estimated to be $Q_r\approx 9800$, which is consistent with the plot. 
We see that the phase noise stays
above the incoming noise $s_0$ while the amplitude noise remains below this value signifying squeezing. They both
approach $s_0$ as the frequency deviates sufficiently beyond the resonator bandwidth (see also Fig. \ref{fig:noises2}).
For $\omega\gg\omega_{\rm roll}$  we find $s^{\rm ph}_{\rm exc}(\omega)\sim(\omega/\omega_0)^{-2}$ and
$-s^{\rm amp}_{\rm exc}(\omega)\sim(\omega/\omega_0)^{-2.9}$.

In Fig. \ref{fig:powerdep}(b) we plot the loss tangent, $Q_{\rm int}^{-1}$ for the pump frequency on resonance and $N=100$. 
This was found numerically by fitting the transmission amplitude to a circuit model as a function of 
$[V(1+Q_{\rm ext}/Q_{\rm int})]^2/Z\approx V^2/Z$. Here, $V$ is the voltage across capacitor $C$ 
(as shown in Fig. \ref{fig:circuit}) and $Q_{\rm ext}=2(C+C_c)/(\omega_0ZC_c^2)$ is the external quality factor. 
The above approximation holds because our resonator is over-coupled. At low amplitude the loss is constant, and at high amplitude 
the slope is $\approx -1$ as the spins are saturated. This is in agreement with the semi-classical theory for a single TLS type. 
For the standard tunneling model 
distribution the density of defects follows $P(\Delta_A,\Delta_0)=P_0/\Delta_0$ and the superposition of the loss from different TLSs 
give a slope of $-1/2$ in the high power regime \cite{schickfus}, as observed in amorphous films. 

In the same figure the excess phase noise $s_{\rm exc}(\omega=0)$ is plotted. In the same high power regime as the loss tangent, 
the partially saturated TLSs exhibit phase noise with a slope of $-1$, which disagrees with the power dependence observed
in experiment \cite{thesis}. This suggests that a correct description for the phase noise power dependence necessitates the
inclusion of intrinsic stochastic fluctuations of the TLSs. 
Indeed, intrinsic TLS noise causes dielectric constant fluctuations and can lead to observable noise \cite{gaoetal2}.
We reiterate, however, that squeezing phenomenon itself is present regardless of deterministic or stochastic nature of TLS dynamics.

In conclusion, we have developed a theoretical framework applicable to circuits containing transmission lines, lumped circuit elements, 
and TLS defects, building upon previous work in quantum optics. It serves as a first step toward a quantitative theory for noise in
these circuits. We found that quadrature-independent incident noise is generally squeezed once transmitted through a lumped 
circuit containing TLSs even when the latter evolve deterministically. Extensions of the model could allow for further 
quantitative results on noise due to TLSs, including the treatment of the standard tunneling model distribution of TLSs
and incorporating intrinsic TLS fluctuations.

\textit{Acknowlegments}: S. T. thanks Lev S. Bishop for discussions. S. T. and V. G. were supported
by the Intelligence Advanced Research Projects Activity (IARPA) through the US army Research Office award
W911NF-09-1-0351. 


\end{document}